# ON REGRESSION ADJUSTMENTS IN EXPERIMENTS WITH SEVERAL TREATMENTS

BY DAVID A. FREEDMAN

*University of California, Berkeley*

Regression adjustments are often made to experimental data. Since randomization does not justify the models, bias is likely; nor are the usual variance calculations to be trusted. Here, we evaluate regression adjustments using Neyman's nonparametric model. Previous results are generalized, and more intuitive proofs are given. A bias term is isolated, and conditions are given for unbiased estimation in finite samples.

**1. Introduction.** Data from randomized controlled experiments (including clinical trials) are often analyzed using regression models and the like. The behavior of the estimates can be calibrated using the nonparametric model in Neyman (1923), where each subject has potential responses to several possible treatments. Only one response can be observed, according to the subject's assignment; the other potential responses must then remain unobserved. Covariates are measured for each subject and may be entered into the regression, perhaps with the hope of improving precision by adjusting the data to compensate for minor imbalances in the assignment groups.

As discussed in Freedman (2006, 2007), randomization does not justify the regression model, so that bias can be expected, and the usual formulas do not give the right variances. Moreover, regression need not improve precision. Here, we extend some of those results, with proofs that are more intuitive. We study asymptotics, isolate a bias term of order $1/n$, and give some special conditions under which the multiple regression estimator is unbiased in finite samples.

What is the source of the bias when regression models are applied to experimental data? In brief, the regression model assumes linear additive effects. Given the assignments, the response is taken to be a linear combination of treatment dummies and covariates, with an additive random error;

coefficients are assumed to be constant across subjects. The Neyman model makes no assumptions about linearity and additivity. If we write the expected response given the assignments as a linear combination of treatment dummies, coefficients will vary across subjects. That is the source of the bias (algebraic details are given below).

To put this more starkly, in the Neyman model, inferences are based on the random assignment to the several treatments. Indeed, the only stochastic element in the model *is* the randomization. With regression, inferences are made conditional on the assignments. The stochastic element is the error term, and the inferences depend on assumptions about that error term. Those assumptions are not justified by randomization. The breakdown in assumptions explains why regression comes up short when calibrated against the Neyman model.

For simplicity, we consider three treatments and one covariate, the main difficulty in handling more variables being the notational overhead. There is a finite population of $n$ subjects, indexed by $i = 1, \ldots, n$. Defined on this population are four variables $a, b, c, z$. The value of $a$ at $i$ is $a_i$, and so forth. These are fixed real numbers. We consider three possible treatments, $A, B, C$. If, for instance, $i$ is assigned to treatment $A$, we observe the response $a_i$, but do not observe $b_i$ or $c_i$.

The population averages are the parameters of interest here:

$$(1) \qquad \overline{a} = \frac{1}{n} \sum_{i=1}^{n} a_i, \qquad \overline{b} = \frac{1}{n} \sum_{i=1}^{n} b_i, \qquad \overline{c} = \frac{1}{n} \sum_{i=1}^{n} c_i.$$

For example, $\overline{a}$ is the average response if all subjects are assigned to $A$. This could be measured directly, at the expense of losing all information about $\overline{b}$ and $\overline{c}$. To estimate all three parameters, we divide the population at random into three sets $A, B, C$, of fixed sizes $n_A$, $n_B$, $n_C$. If $i \in A$, then $i$ receives treatment $A$; likewise for $B$ and $C$. We now have a simple model for a clinical trial. As a matter of notation, $A$ stands for a random set as well as a treatment.

Let $U, V, W$ be dummy variables for the sets. For instance, $U_i = 1$ if $i \in A$ and $U_i = 0$ otherwise. In particular, $\sum_i U_i = n_A$, and so forth. Let $x_A$ be the average of $x$ over $A$, namely,

$$(2) \qquad x_A = \frac{1}{n_A} \sum_{i \in A} x_i.$$

Plainly, $a_A = \sum_{i \in A} a_i / n_A$ is an unbiased estimator, called the "ITT estimator," for $\overline{a}$. Likewise for $B$ and $C$. "ITT" stands for intention-to-treat. The idea, of course, is that the sample average is a good estimator for the population average. The intention-to-treat principle goes back to Bradford Hill (1961); for additional discussion, see Freedman (2006). There is at least one flaw in the notation: $x_A$ is a random variable, being the average of $x$ over the random set $A$. By contrast, $n_A$ is a fixed quantity, being the number of elements in $A$.

In the Neyman model, the observed response for subject $i = 1, \ldots, n$ is

$$(3) \qquad Y_i = a_i U_i + b_i V_i + c_i W_i,$$

because $a, b, c$ code the responses to the treatments. If, for instance, $i$ is assigned to $A$, the response is $a_i$. Furthermore, $U_i = 1$ and $V_i = W_i = 0$, so $Y_i = a_i$. In this circumstance, $b_i$ and $c_i$ would not be observable.

We come now to multiple regression. The variable $z$ is a covariate. It is observed for every subject, and is unaffected by assignment. Applied workers often estimate the parameters in (1) by a multiple regression of $Y$ on $U, V, W, z$. This is the multiple regression estimator whose properties are to be studied. The idea seems to be that estimates are improved by adjusting for random imbalance in assignments.

The standard regression model assumes linear additive effects, so that

$$(4) \qquad E(Y_i | U, V, W, z) = \beta_1 U_i + \beta_2 V_i + \beta_3 W_i + \beta_4 z_i,$$

where $\beta$ is constant across subjects. However, the Neyman model makes no assumptions about linearity or additivity. As a result, $E(Y_i | U, V, W, z)$ is given by the right-hand side of (3), with coefficients that vary across subjects. The variation in the coefficients contradicts the basic assumption needed to prove that regression estimates are unbiased [Freedman (2005), page 43]. The variation in the coefficients is the source of the bias.

Analysts who fit (4) to data from a randomized controlled experiment seem to think of $\hat{\beta}_1$ as estimating the effect of treatment $A$, namely, $\overline{a}$ in (1). Likewise, $\hat{\beta}_3 - \hat{\beta}_1$ is used to estimate $\overline{c} - \overline{a}$, the differential effect of treatment $C$ versus $A$. Similar considerations apply to other effects. However, these estimators suffer from bias and other problems to be explored below.

We turn for a moment to combinatorics. Proposition 1 is a well-known result. (All proofs are deferred to the Appendix at the end of the article.)

PROPOSITION 1. *Let $\tilde{p}_S = n_S / n$ for $S = A, B$ or $C$.*

(i) $E(x_A) = \overline{x}$.
(ii) $\operatorname{var}(x_A) = \frac{1}{n-1} \frac{1 - \tilde{p}_A}{\tilde{p}_A} \operatorname{var}(x)$.
(iii) $\operatorname{cov}(x_A, y_A) = \frac{1}{n-1} \frac{1 - \tilde{p}_A}{\tilde{p}_A} \operatorname{cov}(x, y)$.
(iv) $\operatorname{cov}(x_A, y_B) = -\frac{1}{n-1} \operatorname{cov}(x, y)$.

Here, $x, y = a, b, c$ or $z$. Likewise, $A$ in (i)–(iii) may be replaced by $B$ or $C$. And $A, B$ in (iv) may be replaced by any other distinct pair of sets. By

$\mathrm{cov}(x, y)$, for example, we mean

$$\frac{1}{n} \sum_{i=1}^{n} (x_i - \overline{x})(y_i - \overline{y}).$$

Curiously, the result in (iv) does not depend on the fractions of subjects allocated to the three sets. We can take $x = z$ and $y = z$. For instance,

$$\mathrm{cov}(z_A, z_B) = -\frac{1}{n-1} \mathrm{var}(z).$$

The finite-sample multivariate CLT in Theorem 1 below is a minor variation on results in Höglund (1978). The theorem will be used to prove the asymptotic normality of the multiple regression estimator. There are several regularity conditions for the theorem.

CONDITION #1.   There is an a priori bound on fourth moments. For all $n = 1, 2, \ldots$ and $x = a, b, c$ or $z$,

$$(5) \qquad \frac{1}{n} \sum_{i=1}^{n} |x_i|^4 < L < \infty.$$

CONDITION #2.   The first- and second-order moments, including mixed moments, converge to finite limits, and asymptotic variances are positive. For instance,

$$(6) \qquad \frac{1}{n} \sum_{i=1}^{n} a_i \to \langle a \rangle$$

and

$$(7) \qquad \frac{1}{n} \sum_{i=1}^{n} a_i^2 \to \langle a^2 \rangle, \qquad \frac{1}{n} \sum_{i=1}^{n} a_i b_i \to \langle ab \rangle,$$

with

$$(8) \qquad \langle a^2 \rangle > \langle a \rangle^2;$$

likewise for the other variables and pairs of variables. Here, $\langle a \rangle$ and so forth merely denote finite limits. We take $\langle a^2 \rangle$ and $\langle a, a \rangle$ as synonymous. In present notation, $\langle a \rangle$ is the limit of $\overline{a}$, the latter being the average of $a$ over the population of size $n$; see (1).

CONDITION #3.   We assume groups are of order $n$ in size, that is,

$$(9) \qquad \tilde{p}_A = n_A/n \to p_A > 0, \qquad \tilde{p}_B = n_B/n \to p_B > 0,$$

$$\tilde{p}_C = n_C/n \to p_C > 0,$$

where $p_A + p_B + p_C = 1$. Notice that $\tilde{p}_A$, for instance, is the fraction of subjects assigned to $A$ at stage $n$; the limit as $n$ increases is $p_A$.

Condition #4. The variables $a, b, c, z$ have mean 0:

$$(10) \qquad \frac{1}{n}\sum_{i=1}^{n} x_i = 0, \qquad \text{where } x = a, b, c, z.$$

Condition #4 is a normalization for Theorem 1. Without it, some centering would be needed.

Theorem 1 (The CLT). *Under Conditions #1–#4, the joint distribution of the 12-vector*

$$\sqrt{n}(a_A, a_B, a_C \ldots, z_C)$$

*is asymptotically normal, with parameters given by the limits below:*

(i) $E(\sqrt{n}\,x_A) = 0$;
(ii) $\text{var}(\sqrt{n}\,x_A) \to \langle x^2\rangle(1 - p_A)/p_A$;
(iii) $\text{cov}(\sqrt{n}\,x_A, \sqrt{n}\,y_A) \to \langle x, y\rangle(1 - p_A)/p_A$;
(iv) $\text{cov}(\sqrt{n}\,x_A, \sqrt{n}\,y_B) \to -\langle x, y\rangle$.

Here, $x, y = a, b, c$ or $z$. Likewise, $A$ in (i)–(iii) may be replaced by $B$ or $C$. And $A, B$ in (iv) may be replaced by any other distinct pair of sets. The theorem asserts, among other things, that the limiting first- and second-order moments coincide with the moments of the asymptotic distribution, which is safe due to the bound on fourth moments. (As noted above, proofs are deferred to a Technical Appendix at the end of the article.)

Example 1. Suppose we wish to estimate the effect of $C$ relative to $A$, that is, $\overline{c} - \overline{a}$. The ITT estimator is $Y_C - Y_A = c_C - a_A$, where the equality follows from (3). As before, $Y_C = \sum_{i \in C} Y_i/n_C = \sum_{i \in C} c_i/n_C$. The estimator $Y_C - Y_A$ is unbiased by Proposition 1, and its exact variance is

$$\frac{1}{n-1}\left[\frac{1 - \tilde{p}_A}{\tilde{p}_A}\text{var}(a) + \frac{1 - \tilde{p}_C}{\tilde{p}_C}\text{var}(c) + 2\,\text{cov}(a, c)\right].$$

By contrast, the multiple regression estimator would be obtained by fitting (4) to the data, and computing $\hat{\Delta} = \hat{\beta}_3 - \hat{\beta}_1$. The asymptotic bias and variance of this estimator will be determined in Theorem 2 below. The performance of the two estimators will be compared in Theorem 4.

## 2. Asymptotics for multiple regression estimators.

In this section we state a theorem that describes the asymptotic behavior of the multiple regression estimator applied to experimental data: there is a random term of order $1/\sqrt{n}$ and a bias term of order $1/n$. As noted above, we have three treatments and one covariate $z$. The treatment groups are $A, B, C$, with dummies $U, V, W$. The covariate is $z$. If $i$ is assigned to $A$, we observe the response $a_i$, whereas $b_i, c_i$ remain unobservable. Likewise for $B, C$. The covariate $z_i$ is always observed, and is unaffected by assignment. The response variable $Y$ is given by (3). In Theorem 1, most of the random variables—like $a_B$ or $b_A$—are unobservable. That may affect the applications, but not the mathematics. Arguments below involve only observable random variables.

The design matrix for the multiple regression estimator will have $n$ rows and four columns, namely, $U, V, W, z$. The estimator is obtained by a regression of $Y$ on $U, V, W, z$, the first three coefficients estimating the effects of $A, B, C$, respectively. Let $\hat{\beta}_{\text{MR}}$ be the multiple regression estimator for the effects of $A, B, C$. Thus, $\hat{\beta}_{\text{MR}}$ is a $3 \times 1$-vector.

We normalize $z$ to have mean 0 and variance 1:

$$(11) \qquad \frac{1}{n} \sum_{i=1}^{n} z_i = 0, \qquad \frac{1}{n} \sum_{i=1}^{n} z_i^2 = 1.$$

The mean-zero condition on $z$ overlaps Condition #4, and is needed for Theorem 2. There is no intercept in our regression model; without the mean-zero condition, the mean of $z$ is liable to confound the effect estimates. See the Appendix for details. (In the alternative, we can drop one of the dummies and put an intercept into the regression—although we would now be estimating effect differences rather than effects.) The condition on the mean of $z^2$ merely sets the scale.

Recall that $\tilde{p}_A$ is the fraction of subjects assigned to treatment $A$. Let

$$(12) \qquad \tilde{Q} = \tilde{p}_A \overline{az} + \tilde{p}_B \overline{bz} + \tilde{p}_C \overline{cz}$$

and

$$(13) \qquad Q = p_A \langle az \rangle + p_B \langle bz \rangle + p_C \langle cz \rangle.$$

Here, for instance, $\overline{az} = \sum_{i=1}^{n} a_i z_i / n$ is the average over the study population. By Condition #2, as the population size grows, $\overline{az} = \sum_{i=1}^{n} a_i z_i / n \to \langle az \rangle$; likewise for $b$ and $c$. Thus,

$$(14) \qquad \tilde{Q} \to Q.$$

The quantities $\tilde{Q}$ and $Q$ are needed for the next theorem, which demonstrates asymptotic normality and isolates the bias term. To state the theorem, recall that $\hat{\beta}_{\text{MR}}$ is the multiple regression estimator for the three effects. The estimand is

$$(15) \qquad \beta = (\overline{a}, \overline{b}, \overline{c})',$$

where $\overline{a}, \overline{b}, \overline{c}$ are defined in (1). Define the $3 \times 3$ matrix $\Sigma$ as follows:

$$\Sigma_{11} = \frac{1 - p_A}{p_A} \lim \text{var}(a - Qz),$$
$$(16)$$
$$\Sigma_{12} = -\lim \text{cov}(a - Qz, b - Qz),$$

and so forth. The limits are taken as the population size $n \to \infty$, and exist by Condition #2. Let

$$(17) \qquad \zeta_n = \sqrt{n}(a_A - \tilde{Q}z_A, b_B - \tilde{Q}z_B, c_C - \tilde{Q}z_C)'.$$

This turns out to be the lead random element in $\hat{\beta}_{\mathrm{MR}} - \beta$. The asymptotic variance–covariance matrix of $\zeta_n$ is $\Sigma$, by (14) and Theorem 1. For the bias term, let

$$(18) \qquad K_A = \mathrm{cov}(az, z) - \tilde{p}_A \, \mathrm{cov}(az, z) - \tilde{p}_B \, \mathrm{cov}(bz, z) - \tilde{p}_C \, \mathrm{cov}(cz, z),$$

and likewise for $K_B, K_C$.

THEOREM 2. *Assume Conditions #1–#3, not Condition #4, and (11). Define $\zeta_n$ by (17), and $K_S$ by (18) for $S = A, B, C$. Then $E(\zeta_n) = 0$ and $\zeta_n$ is asymptotically $N(0, \Sigma)$. Moreover,*

$$(19) \qquad \hat{\beta}_{\mathrm{MR}} - \beta = \zeta_n/\sqrt{n} - K/n + \rho_n,$$

*where $K = (K_A, K_B, K_C)'$ and $\rho_n = O(1/n^{3/2})$ in probability.*

REMARKS. (i) If $K = 0$, the bias term will be $O(1/n^{3/2})$ or smaller.

(ii) What are the implications for practice? In the usual linear model, $\hat{\beta}$ is unbiased given $X$. With experimental data and the Neyman model, given the assignment, results are deterministic. At best, we will get unbiasedness on average, over all assignments. Under special circumstances (Theorems 5 and 6 below), that happens. Generally, however, the multiple regression estimator will be biased. See Example 5. The bias decreases as sample size increases.

(iii) Turn now to random error in $\hat{\beta}$. This is of order $1/\sqrt{n}$, both for the ITT estimator and for the multiple regression estimator. However, the asymptotic variances differ. The multiple regression estimator can be more efficient than the ITT estimator—or less efficient—and the difference persists even for large samples. See Examples 3 and 4 below.

## 3. Asymptotic nominal variances.

"Nominal" variances are computed by the usual regression formulae, but are likely to be wrong since the usual assumptions do not hold. We sketch the asymptotics here, under the conditions of Theorem 2. Recall that the design matrix $X$ is $n \times 4$, the columns being $U, V, W, z$. The response variable is $Y$. The nominal covariance matrix is then

$$(20) \qquad \Sigma_{\mathrm{nom}} = \hat{\sigma}^2 (X'X)^{-1},$$

where $\hat{\sigma}^2$ is the sum of the squared residuals, normalized by the degrees of freedom $(n - 4)$. Recall $Q$ from (13). Let

$$(21) \qquad \sigma^2 = \lim_{n \to \infty} [\tilde{p}_A \, \mathrm{var}(a) + \tilde{p}_B \, \mathrm{var}(b) + \tilde{p}_C \, \mathrm{var}(c)] - Q^2,$$

where the limit exists by Conditions #2 and #3. Let

$$(22) \qquad D = \begin{pmatrix} p_A & 0 & 0 & 0 \\ 0 & p_B & 0 & 0 \\ 0 & 0 & p_C & 0 \\ 0 & 0 & 0 & 1 \end{pmatrix}.$$

THEOREM 3. *Assume Conditions #1–#3, not Condition #4, and (11). Define $\sigma^2$ by (21) and $D$ by (22). In probability,*

(i) $X'X/n \to D$,

(ii) $\hat{\sigma}^2 \to \sigma^2$,

(iii) $n\Sigma_{\text{nom}} \to \sigma^2 D^{-1}$.

What are the implications for practice? The upper left $3 \times 3$ block of $\sigma^2 D^{-1}$ will generally differ from $\Sigma$ in Theorem 2, so the usual regression standard errors—computed for experimental data—can be quite misleading. This difficulty does not go away for large samples. What explains the breakdown? In brief, the multiple regression assumes (i) the expectation of the response given the assignment variables and the covariates is linear, with coefficients that are constant across subjects; and (ii) the conditional variance of the response is constant across subjects. In the Neyman model, (i) is wrong as noted earlier. Moreover, given the assignments, there is no variance left in the responses.

More technically, variances in the Neyman model are (necessarily) computed across the assignments, for it is the assignments that are the random elements in the model. With regression, variances are computed conditionally on the assignments, from an error term assumed to be IID across subjects, and independent of the assignment variables as well as the covariates. These assumptions do not follow from the randomization, explaining why the usual formulas break down. For additional discussion, see Freedman (2007).

An example may clarify the issues. Write $\text{cov}_\infty$ for limiting covariances, for example,

$$\text{cov}_\infty(a, z) = \lim \text{cov}(a, z) = \langle az \rangle - \langle a \rangle \langle z \rangle = \langle az \rangle$$

because $\langle z \rangle = 0$ by (11); similarly for variances. See Condition #2.

EXAMPLE 2. Consider estimating the effect of $C$ relative to $A$, so the parameter of interest is $\overline{c} - \overline{a}$. By way of simplification, suppose $Q = 0$. Let $\hat{\Delta}$ be the multiple regression estimator for the effect difference. By Theorem 3, the nominal variance of $\hat{\Delta}$ is essentially $1/n$ times

$$\left(1 + \frac{p_A}{p_C}\right) \text{var}_\infty(a) + \left(1 + \frac{p_C}{p_A}\right) \text{var}_\infty(c) + \left(\frac{1}{p_A} + \frac{1}{p_C}\right) p_B \, \text{var}_\infty(b).$$

By Theorem 2, however, the true asymptotic variance of $\hat{\Delta}$ is $1/n$ times

$$\left(\frac{1}{p_A} - 1\right) \text{var}_\infty(a) + \left(\frac{1}{p_C} - 1\right) \text{var}_\infty(c) + 2\,\text{cov}_\infty(a, c).$$

For instance, we can take the asymptotic variance-covariance matrix of $a, b, c, z$ to be the $4 \times 4$ identity matrix, with $p_A = p_C = 1/4$ so $p_B = 1/2$. The true asymptotic variance of $\hat{\Delta}$ is $6/n$. The nominal asymptotic variance is $8/n$ and is too big. On the other hand, if we change $\text{var}_\infty(b)$ to $1/4$, the true asymptotic variance is still $6/n$; the nominal asymptotic variance drops to $5/n$ and is too small.

**4. The gain from adjustment.** Does adjustment improve precision? The answer is sometimes.

THEOREM 4. *Assume Conditions #1–#3, not Condition #4, and (11). Consider estimating the effect of $C$ relative to $A$, so the parameter of interest is $\overline{c} - \overline{a}$. If we compare the multiple regression estimator to the ITT estimator, the asymptotic gain in variance is $\Gamma/(np_Ap_C)$, where*

$$(23) \qquad \Gamma = 2Q[p_C\langle az\rangle + p_A\langle cz\rangle] - Q^2[p_A + p_C],$$

*with $Q$ defined by (13). Adjustment therefore helps asymptotic precision if $\Gamma > 0$, but hurts if $\Gamma < 0$.*

The next two examples are set up like Example 2, with $\text{cov}_\infty$ for limiting covariances. We say the design is *balanced* if $n$ is a multiple of 3 and $n_A = n_B = n_C = n/3$. We say that effects are *additive* if $b_i - a_i$ is constant over $i$ and likewise for $c_i - a_i$. With additive effects, $\text{var}_\infty(a) = \text{var}_\infty(b) = \text{var}_\infty(c)$; write $v$ for the common value. Similarly, $\text{cov}_\infty(a, z) = \text{cov}_\infty(b, z) = \text{cov}_\infty(c, z) = Q = \rho\sqrt{v}$, where $\rho$ is the asymptotic correlation between $a$ and $z$, or $b$ and $z$, or $c$ and $z$.

EXAMPLE 3. Suppose effects are additive. Then $\text{cov}_\infty(a, z) = \text{cov}_\infty(b, z) = \text{cov}_\infty(c, z) = Q$ and $\Gamma = Q^2(p_A + p_C) \geq 0$. The asymptotic gain from adjustment will be positive if $\text{cov}_\infty(a, z) \neq 0$.

EXAMPLE 4. Suppose the design is balanced, so $p_A = p_B = p_C = 1/3$. Then $3Q = \text{cov}_\infty(a, z) + \text{cov}_\infty(b, z) + \text{cov}_\infty(c, z)$. Consequently, $3\Gamma/2 = Q[2Q - \text{cov}_\infty(b, z)]$. Let $z = a + b + c$. Choose $a, b, c$ so that $\text{var}_\infty(z) = 1$ and $\text{cov}_\infty(a, b) = \text{cov}_\infty(a, c) = \text{cov}_\infty(b, c) = 0$. In particular, $Q = 1/3$. Now $2Q - \text{cov}_\infty(b, z) = 2/3 - \text{var}_\infty(b)$. The asymptotic gain from adjustment will be negative if $\text{var}_\infty(b) > 2/3$.

Example 3 indicates one motivation for adjustment: if effects are nearly additive, adjustment is likely to help. However, Example 4 shows that even in a balanced design, the "gain" from adjustment can be negative—if there are subject-by-treatment interactions. More complicated and realistic examples can no doubt be constructed.

**5. Finite-sample results.** This section gives conditions under which the multiple regression estimator will be exactly unbiased in finite samples. Arguments are from symmetry. As before, the design is balanced if $n$ is a multiple of 3 and $n_A = n_B = n_C = n/3$; effects are additive if $b_i - a_i$ is constant over $i$ and likewise for $c_i - a_i$. Then $a_i - \overline{a} = b_i - \overline{b} = c_i - \overline{c} = \delta_i$, say, for all $i$. Note that $\sum_i \delta_i = 0$.

THEOREM 5.    *If (11) holds, the design is balanced, and effects are additive, then the multiple regression estimator is unbiased.*

Examples show that the balance condition is needed in Theorem 5: additivity is not enough. Likewise, if the balance condition holds but there is nonadditivity, the multiple regression estimator will usually be biased. We illustrate the first point.

EXAMPLE 5.    Consider a miniature trial with 6 subjects. Responses $a, b, c$ to treatments $A, B, C$ are shown in Table 1, along with the covariate $z$. Notice that $b - a = 1$ and $c - a = 2$. Thus, effects are additive. We assign one subject at random to $A$, one to $B$, and the remaining four to $C$. There are $6 \times 5/2 = 15$ assignments. For each assignment, we build up the $6 \times 4$ design matrix (one column for each treatment dummy and one column for $z$); we compute the response variable from Table 1 above, and then the multiple regression estimator. Finally, we average the results across the 15 assignments, as shown in Table 2. The average gives the expected value of the multiple regression estimator, because the average is taken across all possible designs. "Truth" is determined from the parameters in Table 1. Calculations are exact, within the limits of rounding error; no simulations are involved.

For instance, the average coefficient for the $A$ dummy is 3.3825. However, from Table 1, the average effect of $A$ is $\overline{a} = 1.3333$. The difference is bias. Consider next the differential effect of $B$ versus $A$. On average, this is estimated by multiple regression as $1.9965 - 3.3825 = -1.3860$. From Table 1, truth is $+1$. Again, this reflects bias in the multiple regression estimator. With a larger trial, of course, the bias would be smaller; see Theorem 2. Theorem 5 does not apply because the design is unbalanced.

TABLE 1
*Parameter values*

| $a$ | $b$ | $c$ | $z$ |
|-----|-----|-----|-----|
| 0 | 1 | 2 | 0 |
| 0 | 1 | 2 | 0 |
| 0 | 1 | 2 | 0 |
| 2 | 3 | 4 | −2 |
| 2 | 3 | 4 | −2 |
| 4 | 5 | 6 | 4 |

For the next theorem, consider the possible values $v$ of $z$. Let $n_v$ be the number of $i$ with $z_i = v$. The average of $a_i$ given $z_i = v$ is

$$\frac{1}{n_v} \sum_{\{i : z_i = v\}} a_i.$$

Suppose this is constant across $v$'s, as is $\sum_{\{i : z_i = v\}} b_i / n_v$, $\sum_{\{i : z_i = v\}} c_i / n_v$. The common values must be $\overline{a}$, $\overline{b}$, $\overline{c}$, respectively. We call this *conditional constancy*. No condition is imposed on $z$, and the design need not be balanced. (Conditional constancy is violated in Example 5, as one sees by looking at the parameter values in Table 1.)

THEOREM 6. *With conditional constancy, the multiple regression estimator is unbiased.*

REMARKS. (i) In the usual regression model, $Y = X\beta + \epsilon$ with $E(\epsilon | X) = 0$. The multiple regression estimator is then conditionally unbiased. In Theorems 5 and 6, the estimator is conditionally biased, although the bias averages out to 0 across permutations. In Theorem 5, for instance, the conditional bias is $(X'X)^{-1}X'\delta$. Across permutations, the bias averages out to 0. The proof is a little tricky (see the Technical Appendix below). The $\delta$ is fixed, as explained before the theorem; it is $X$ that varies from one permutation to another; the conditional bias is a nonlinear function of $X$. This is all quite different from the usual regression arguments.

TABLE 2
*Average multiple regression estimates versus truth*

|   | **Ave MR** | **Truth** |
|---|-----------|-----------|
| $A$ | 3.3825 | 1.3333 |
| $B$ | 1.9965 | 2.3333 |
| $C$ | 2.9053 | 3.3333 |
| $z$ | −0.0105 |   |

(ii) Kempthorne (1952) points to the difference between permutation models and the usual linear regression model; see Chapters 7–8, especially Section 8.7. Also see *Biometrics* vol. 13, no. 3 (1957). Cox (1956) cites Kempthorne, but appears to contradict Theorem 5 above. I am indebted to Joel Middleton for the reference to Cox.

(iii) When specialized to two-group experiments, the formulas in this paper (for, e.g., asymptotic variances) differ in appearance but not in substance from those previously reported [Freedman (2007)].

(iv) Although details have not been checked, the results (and the arguments) in this paper seem to extend easily to any fixed number of treatments, and any fixed number of covariates. Treatment by covariate interactions can probably be accommodated too.

(v) In this paper treatments have two levels: low or high. If a treatment has several levels—for example, low, medium, high—and linearity is assumed in a regression model, inconsistency is likely to be a consequence. Likewise, we view treatments as mutually exclusive: if subject $i$ is assigned to group $A$, then $i$ cannot also turn up in group $B$. If multiple treatments are applied to the same subject in order to determine joint effects, and a regression model assumes additive or multiplicative effects, inconsistency is again likely.

(vi) The theory developed here applies equally well to 0–1 valued responses. With 0–1 variables, it may seem more natural to use logit or probit models to adjust the data. However, such models are not justified by randomization—any more than the linear model. Preliminary calculations suggest that if adjustments are to be made, linear regression may be a safer choice. For instance, the conventional logit estimator for the odds ratio may be severely biased. On the other hand, a consistent estimator can be based on estimated probabilities in the logit model. For discussion, see Freedman (2008).

(vii) The theory developed here can probably be extended to more complex designs (like blocking) and more complex estimators (like two-stage least squares), but the work remains to be done.

(viii) Victora, Habicht and Bryce (2004) favor adjustment. However, they do not address the sort of issues raised here, nor are they entirely clear about whether inferences are to be made on average across assignments, or conditional on assignment. In the latter case, inferences might be strongly model-dependent.

(ix) Models are used to adjust data from large randomized controlled experiments in, for example, Cook et al. (2007), Gertler (2004), Chattopadhyay and Duflo (2004) and Rossouw et al. (2002). Cook et al. report on long-term followup of subjects in experiments where salt intake was restricted; conclusions are dependent on the models used to analyze the data. By contrast, the results in Rossouw et al. for hormone replacement therapy do not depend very much on the modeling.

**6. Recommendations for practice.** Altman et al. ([2001](#)) document persistent failures in the reporting of data from clinical trials, and make detailed proposals for improvement. The following recommendations are complementary:

(i) As is usual, measures of balance between the assigned-to-treatment group and the assigned-to-control group should be reported.

(ii) After that should come a simple intention-to-treat analysis, comparing rates (or averages and SDs) of outcomes among those assigned to treatment and those assigned to the control group.

(iii) Crossover should be discussed, and deviations from protocol.

(iv) Subgroup analyses should be reported, and corrections for crossover if that is to be attempted. Analysis by treatment received requires special justification, and so does per protocol analysis. (The first compares those who receive treatment with those who do not, regardless of assignment; the second censors subjects who cross over from one arm of the trial to the other, e.g., they are assigned to control but insist on treatment.) Complications are discussed in Freedman ([2006](#)).

(v) Regression estimates (including logistic regression and proportional hazards) should be deferred until rates and averages have been presented. If regression estimates differ from simple intention-to-treat results, and reliance is placed on the models, that needs to be explained. As indicated above, the usual models are not justified by randomization, and simpler estimators may be more robust.

## TECHNICAL APPENDIX

The [Appendix](#) provides technical underpinnings for the theorems discussed above.

PROOF OF PROPOSITION [1](#). We prove only claim (iv). Plainly, $E(U_i V_j) = 0$ if $i = j$, since $i$ cannot be assigned both to $A$ and to $B$. Furthermore,

$$E(U_i V_j) = P(U_i = 1 \,\&\, V_j = 1) = \frac{n_A}{n}\frac{n_B}{n-1}$$

if $i \neq j$. This is clear if $i = 1$ and $j = 2$; but permuting indices will not change the joint distribution of assignment dummies. We may assume without loss of generality that $\overline{x} = \overline{y} = 0$. Now

$$\mathrm{cov}(x_A, y_B) = \frac{1}{n_A}\frac{1}{n_B}\sum_{i \neq j} E(U_i V_j x_i y_j)$$

$$= \frac{1}{n(n-1)}\sum_{i \neq j} x_i y_j$$

$$= \frac{1}{n(n-1)} \left( \sum_i x_i \sum_j y_j - \sum_i x_i y_i \right)$$

$$= -\frac{1}{n(n-1)} \sum_i x_i y_i = -\frac{1}{n-1} \operatorname{cov}(x, y)$$

as required, where $i, j = 1, \ldots, n$. $\quad\square$

PROOF OF THEOREM 1. The theorem can be proved by appealing to Höglund (1978) and computing conditional distributions. Another starting point is Hoeffding (1951), with suitable choices for the matrix from which summands are drawn. With either approach, the usual linear-combinations trick can be used to reduce dimensionality. In view of (9), the limiting distribution satisfies three linear constraints.

A formal proof is omitted, but we sketch the argument for one case, starting from Theorem 3 in Hoeffding (1951). Let $\alpha, \beta, \gamma$ be three constants. Let $M$ be an $n \times n$ matrix, with

$$M_{ij} = \begin{cases} \alpha a_j, & \text{for } i = 1, \ldots, n_A, \\ \beta b_j, & \text{for } i = n_A + 1, \ldots, n_A + n_B, \\ \gamma c_j, & \text{for } i = n_A + n_B + 1, \ldots, n. \end{cases}$$

Pick one $j$ at random from each row, without replacement (interpretation: if $j$ is picked from row $i = 1, \ldots, n_A$, subject $j$ goes into treatment group $A$). According to Hoeffding's theorem, the sum of the corresponding matrix entries will be approximately normal. So the law of $\sqrt{n}(a_A, b_B, c_C)$ tends to multivariate normal. Theorem 1 in Hoeffding's paper will help get the regularity conditions in his Theorem 3 from Conditions #1–#4 above. $\quad\square$

Let $X$ be an $n \times p$ matrix of rank $p \leq n$. Let $Y$ be an $n \times 1$ vector. The multiple regression estimator computed from $Y$ is $\hat{\beta}_Y = (X'X)^{-1} X'Y$. Let $\theta$ be a $p \times 1$ vector. The "invariance lemma" is a purely arithmetic result; the well-known proof is omitted.

LEMMA A.1. *The invariance lemma.* $\hat{\beta}_{Y+X\theta} = \hat{\beta}_Y + \theta$.

The multiple-regression estimator for Theorem 2 may be computed as follows. Recall from (2) that $Y_A$ is the average of $Y$ over $A$, that is, $\sum_{i \in A} Y_i / n_A$; likewise for $B, C$. Let

(A1)                $e_i = Y_i - Y_A U_i - Y_B V_i - Y_C W_i,$

which is the residual when $Y$ is regressed on the first three columns of the design matrix. Let

(A2)                $f_i = z_i - z_A U_i - z_B V_i - z_C W_i,$

which is the residual when $z$ is regressed on those columns. Let $\hat{Q}$ be the slope when $e$ is regressed on $f$:

$$(A3) \qquad \hat{Q} = e \cdot f / |f|^2.$$

The next result is standard.

LEMMA A.2. *The multiple regression estimator for the effect of $A$, that is, the first element in $(X'X)^{-1}X'Y$, is*

$$(A4) \qquad Y_A - \hat{Q}z_A$$

*and likewise for $B, C$. The coefficient of $z$ in the regression of $Y$ on $U, V, W, z$ is $\hat{Q}$.*

We turn now to $\hat{Q}$; this is the key technical quantity in the paper, and we develop a more explicit formula for it. Notice that the dummy variables $U, V, W$ are mutually orthogonal. By the usual regression arguments,

$$(A5) \qquad |f|^2 = |z|^2 - n_A(z_A)^2 - n_B(z_B)^2 - n_C(z_C)^2,$$

where $|f|^2 = \sum_{i=1}^n f_i^2$. Recall (3). Check that $Y_A = a_A$, where $a_A = \sum_{i \in A} a_i / n_A$; likewise for $B, C$. Hence,

$$(A6) \qquad e_i = (a_i - a_A)U_i + (b_i - b_B)V_i + (c_i - c_C)W_i,$$

where the residual $e_i$ was defined in (A1). Likewise,

$$(A7) \qquad f_i = (z_i - z_A)U_i + (z_i - z_B)V_i + (z_i - z_C)W_i,$$

where the residual $f_i$ was defined in (A2). Now

$$(A8) \qquad \begin{aligned} e_i f_i &= (a_i - a_A)(z_i - z_A)U_i + (b_i - b_B)(z_i - z_B)V_i \\ &\quad + (c_i - c_C)(z_i - z_C)W_i \end{aligned}$$

and

$$(A9) \qquad \begin{aligned} \sum_{i=1}^n e_i f_i &= n_A[(az)_A - a_A z_A] + n_B[(bz)_B - b_B z_B] \\ &\quad + n_C[(cz)_C - c_C z_C], \end{aligned}$$

where, for instance, $(az)_A = \sum_{i \in A} a_i z_i / n_A$.

Recall that $\tilde{p}_A = n_A/n$ is the fraction of subjects assigned to treatment $A$; likewise for $B$ and $C$. These fractions are deterministic, not random. We can now give a more explicit formula for the $\hat{Q}$ defined in (A3), dividing numerator and denominator by $n$. By (A5) and (A9),

$$(A10) \quad \begin{aligned} \hat{Q} &= N/D, \qquad \text{where} \\ N &= \tilde{p}_A[(az)_A - a_A z_A] + \tilde{p}_B[(bz)_B - b_B z_B] + \tilde{p}_C[(cz)_C - c_C z_C], \\ D &= 1 - \tilde{p}_A(z_A)^2 - \tilde{p}_B(z_B)^2 - \tilde{p}_C(z_C)^2. \end{aligned}$$

In the formula for $D$, we used (11) to replace $|z|^2/n$ by 1.

The reason $\hat{Q}$ matters is that it relates the multiple regression estimator to the ITT estimator in a fairly simple way. Indeed, by (3) and Lemma A.2,

$$(A11) \qquad \begin{aligned} \hat{\beta}_{\mathrm{MR}} &= (Y_A - \hat{Q}z_A, Y_B - \hat{Q}z_B, Y_C - \hat{Q}z_C)' \\ &= (a_A - \hat{Q}z_A, a_B - \hat{Q}z_B, a_C - \hat{Q}z_C)'. \end{aligned}$$

We must now estimate $\hat{Q}$. In view of (11), Theorem 1 shows that

$$(A12) \qquad (z_A, z_B, z_C) = O(1/\sqrt{n}).$$

(All $O$'s are in probability.) Consequently,

$$(A13) \qquad \text{the denominator } D \text{ of } \hat{Q} \text{ in (A10) is } 1 + O(1/n).$$

Two deterministic approximations to the numerator $N$ were presented in (12)–(13).

PROOF OF THEOREM 2. By Lemma A.1, we may assume $\overline{a} = \overline{b} = \overline{c} = 0$. To see this more sharply, recall (3). Let $\hat{\beta}$ be the result of regressing $Y$ on $U, V, W, z$. Furthermore, let

$$(A14) \qquad Y_i^* = (a_i + a^*)U_i + (b_i + b^*)V_i + (c_i + c^*)W_i.$$

The result of regressing $Y^*$ on $U, V, W, z$ is just $\hat{\beta} + (a^*, b^*, c^*, 0)'$. So the general case of Theorem 2 would follow from the special case. That is why we can, without loss of generality, assume Condition #4. Now

$$(A15) \qquad (a_A, b_B, c_C) = O(1/\sqrt{n}).$$

We use (A10) to evaluate (A11). The denominator of $\hat{Q}$ is essentially 1, that is, the departure from 1 can be swept into the error term $\rho_n$, because the departure from 1 gets multiplied by $(z_A, z_B, z_C)' = O(1/\sqrt{n})$. This is a little delicate, we are estimating down to order $1/n^{3/2}$. The departure of the denominator from 1 is multiplied by $N$, but terms like $a_A z_A$ are $O(1/n)$ and immaterial, while terms like $(az)_A$ are $O(1)$ by Condition #1 and Proposition 1 (or see the discussion of Proposition A.1 below).

For the numerator of $\hat{Q}$, terms like $a_A z_A$ go into $\rho_n$: after multiplication by $(z_A, z_B, z_C)'$, they are $O(1/n^{3/2})$. Recall that $\overline{az} = \sum_{i=1}^n a_i z_i/n$. What's left of the numerator is $\check{Q} + \tilde{Q}$, where

$$(A16) \qquad \check{Q} = \tilde{p}_A(az - \overline{az})_A + \tilde{p}_B(bz - \overline{bz})_B + \tilde{p}_C(cz - \overline{cz})_C.$$

The term $\check{Q}(z_A, z_B, z_C)'$ goes into $\zeta_n$; see (17). The rest of $\zeta_n$ comes from $(a_A, b_B, c_C)$ in (A11). The bias in estimating the effects is therefore

$$(A17) \qquad -E\left\{ \tilde{Q} \begin{pmatrix} z_A \\ z_B \\ z_C \end{pmatrix} \right\}.$$

This can be evaluated by Proposition 1, the relevant variables being $az, bz, cz, z$.
□

ADDITIONAL DETAIL FOR THEOREM 2. We need to show, for instance,

$$\hat{Q} z_A = \tilde{Q} z_A + \check{Q} z_A + O\left(\frac{1}{n^{3/2}}\right).$$

This can be done in three easy steps.

*Step* 1.

$$\frac{N}{D} z_A = N z_A + O\left(\frac{1}{n^{3/2}}\right).$$

Indeed, $N = O(1)$, $D = 1 + O(\frac{1}{n})$, and $z_A = O(\frac{1}{\sqrt{n}})$.

*Step* 2. $N = \tilde{Q} + \check{Q} - R$, where $R = \tilde{p}_A a_A z_A + \tilde{p}_B b_B z_B + \tilde{p}_C c_C z_C$. This is because $\overline{(az)}_A = \overline{az}$, and so forth.

*Step* 3. $R = O(\frac{1}{n})$ so $R z_A = O(\frac{1}{n^{3/2}})$.

REMARKS. (i) As a matter of notation, $\tilde{Q}$ is deterministic but $\check{Q}$ is random. Both are scalar: compare (12) and (A16). The source of the bias is the covariance between $\check{Q}$ and $z_A, z_B, z_C$.

(ii) Suppose we add a constant $k$ to $z$. Instead of (A11), we get $\overline{z} = k$ and $\overline{z^2} = 1 + k^2$. Because $z_A$ and so forth are all shifted by the same amount $k$, the shift does not affect $e, f$ or $\hat{Q}$; see (A1)–(A3). The multiple regression estimator for the effect of $A$ is therefore shifted by $\hat{Q}k$; likewise for $B, C$. This bias does not tend to 0 when sample size grows, but does cancel when estimating differences in effects.

(iii) In applications, we cannot assume the parameters $\overline{a}, \overline{b}, \overline{c}$ are 0—the whole point is to estimate them. The invariance lemma, however, reduces the general case to the more manageable special case, where $\overline{a} = \overline{b} = \overline{c} = 0$, as in the proof of Theorem 2.

(iv) In (19), $K = O(1)$. Indeed, $\overline{z} = 0$, so $\operatorname{cov}(az, z) = \overline{(az)z} = \overline{az^2}$. Now

$$\left|\frac{1}{n}\sum_{i=1}^{n} a_i z_i^2\right| \le \left(\frac{1}{n}\sum_{i=1}^{n} |a_i|^3\right)^{1/3} \left(\frac{1}{n}\sum_{i=1}^{n} |z_i|^3\right)^{2/3}$$

by Hölder's inequality applied to $a$ and $z^2$. Finally, use Condition #1. The same argument can be used for $\operatorname{cov}(bz, z)$ and $\operatorname{cov}(cz, z)$.

Define $\hat{Q}$ as in (A3); recall (A1)–(A2). The residuals from the multiple regression are $e - \hat{Q}f$ by Lemma A.2; according to usual procedures,

$$(A18) \qquad \hat{\sigma}^2 = |e - \hat{Q}f|^2/(n-4).$$

Recall $f$ from (A2), and $\hat{Q}, Q$ from (A3) and (13).

LEMMA A.3.    *Assume Conditions #1-#3, not Condition #4, and (11). Then $|f|^2/n \to 1$ and $\hat{Q} \to Q$. Convergence is in probability.*

PROOF.    The first claim follows from (A5) and (A12); the second, from (A10) and Theorem 1.    □

PROOF OF THEOREM 3.    Let $M$ be the $4 \times 4$ matrix whose diagonal is $\tilde{p}_A, \tilde{p}_B, \tilde{p}_C, 1$; the last row of $M$ is $(z_A, z_B, z_C, 1)$; the last column of $M$ is $(z_A, z_B, z_C, 1)'$. Pad out $M$ with 0's. Plainly, $X'X/n = M$. As before, $\tilde{p}_A = n_A/n$ is deterministic, and $\tilde{p}_A \to p_A$ by (9). But $z_A = O(1/\sqrt{n})$; likewise for $B, C$. This proves (i).

For (ii), $e = e - \hat{Q}f + \hat{Q}f$. But $e - \hat{Q}f \perp f$. So $|e - \hat{Q}f|^2 = |e|^2 - \hat{Q}^2|f|^2$. Then

$$\frac{n-4}{n}\hat{\sigma}^2 = \frac{|e - \hat{Q}f|^2}{n}$$

$$= \frac{|e|^2 - \hat{Q}^2|f|^2}{n}$$

$$= \frac{|Y|^2}{n} - \tilde{p}_A(Y_A)^2 - \tilde{p}_B(Y_B)^2 - \tilde{p}_C(Y_C)^2 - \hat{Q}^2\frac{|f|^2}{n}$$

$$= \frac{|Y|^2}{n} - \tilde{p}_A(a_A)^2 - \tilde{p}_B(b_B)^2 - \tilde{p}_C(c_C)^2 - \hat{Q}^2\frac{|f|^2}{n}$$

by (A1) and (3). Using (3) again, we get

$$(A19) \qquad \frac{|Y|^2}{n} = \tilde{p}_A(a^2)_A + \tilde{p}_B(b^2)_B + \tilde{p}_C(c^2)_C.$$

(Remember, the dummy variables are orthogonal.) So

$$(A20) \qquad \frac{n-4}{n}\hat{\sigma}^2 = \tilde{p}_A[(a^2)_A - (a_A)^2] + \tilde{p}_B[(b^2)_B - (b_B)^2]$$

$$+ \tilde{p}_C[(c^2)_C - (c_C)^2] - \hat{Q}^2\frac{|f|^2}{n}.$$

To evaluate $\lim \hat{\sigma}^2$, we may without loss of generality assume Condition #4, by the invariance lemma. Now $a_A = O(1/\sqrt{n})$ and likewise for $B, C$ by (A15). The terms in (A20) involving $(a_A)^2, (b_B)^2, (c_C)^2$ can therefore be dropped, being $O(1/n)$. Furthermore, $|f|^2/n \to 1$ and $\hat{Q} \to Q$ by Lemma A.3. To complete the proof of (ii), we must show that, in probability,

$$(A21) \qquad (a^2)_A \to \langle a^2 \rangle, \qquad (b^2)_B \to \langle b^2 \rangle, \qquad (c^2)_C \to \langle c^2 \rangle.$$

This follows from Condition #1 and Proposition 1. Given (i) and (ii), claim (iii) is immediate.    □

Proof of Theorem 4. The asymptotic variance of the multiple regression estimator is given by Theorem 2. The variance of the ITT estimator $Y_C - Y_A$ can be worked out exactly, from Proposition 1 (see Example 1). A bit of algebra will now prove Theorem 4. □

Proof of Theorem 5. By the invariance lemma, we may as well assume that $\overline{a} = \overline{b} = \overline{c} = 0$. The ITT estimator is unbiased. By Lemma A.2, the multiple regression estimator differs from the ITT estimator by $\hat{Q}z_A, \hat{Q}z_B, \hat{Q}z_C$. These three random variables sum to 0 by (11) and the balance condition. So their expectations sum to 0. Moreover, the three random variables are exchangeable, so their expectations must be equal. To see the exchangeability more sharply, recall (A1)–(A3). Because there are no interactions, $Y_i = \delta_i$. So

(A27) $$e = \delta - \delta_A U - \delta_B V - \delta_C W$$

by (A1), and

(A28) $$f = z - z_A U - z_B V - z_C W$$

by (A2). These are random $n$-vectors. The joint distribution of

(A29) $$e, f, \hat{Q}, z_A, z_B, z_C$$

does not depend on the labels $A, B, C$: the pairs $(\delta_i, z_i)$ are just being divided into three random groups of equal size. □

The same argument shows that the multiple regression estimator for an effect difference (like $\overline{a} - \overline{c}$) is symmetrically distributed around the true value.

Proof of Theorem 6. By Lemma A.1, we may assume without loss of generality that $\overline{a} = \overline{b} = \overline{c} = 0$. We can assign subjects to $A, B, C$ by randomly permuting $\{1, 2, \ldots, n\}$: the first $n_A$ subjects go into $A$, the next $n_B$ into $B$, and the last $n_C$ into $C$. Freeze the number of $A$'s, $B$'s—and hence $C$'s— within each level of $z$. Consider only the corresponding permutations. Over those permutations, $z_A$ is frozen; likewise for $B, C$. So the denominator of $\hat{Q}$ is frozen: without condition (11), the denominator must be computed from (A5). In the numerator, $z_A, z_B, z_C$ are frozen, while $a_A$ averages out to zero over the permutations of interest; so do $b_B$ and $c_C$. With a little more effort, one also sees that $(az)_A$ averages out to zero, as do $(bz)_B, (cz)_C$. In consequence, $\hat{Q}z_A$ has expectation 0, and likewise for $B, C$. Lemma A.2 completes the argument. □

REMARKS. (i) What if $|f| = 0$ in (A2)–(A3)? Then $z$ is a linear combination of the treatment dummies $U, V, W$; the design matrix $(UVWz)$ is singular, and the multiple regression estimator is ill-defined. This is not a problem for Theorems 2 or 3, being a low-probability event. But it is a problem for Theorems 4 and 5. The easiest course is to assume the problem away, for instance, requiring

(A30)        $z$ is linearly independent of the treatment dummies for every permutation of $\{1, 2, \ldots, n\}$.

Another solution is more interesting: exclude the permutations where $|f| = 0$, and show the multiple regression estimator is conditionally unbiased, that is, has the right average over the remaining permutations.

(ii) All that is needed for Theorems 2–4 is an a priori bound on absolute third moments in Condition #1, rather than fourth moments; third moments are used for the CLT by Höglund (1978). The new awkwardness is in proving results like (A21), but this can be done by familiar truncation arguments. More explicitly, let $x_1, \ldots, x_n$ be real numbers, with

(A31)        $$\frac{1}{n} \sum_{i=1}^{n} |x_i|^\alpha < L.$$

Here, $1 < \alpha < \infty$ and $0 < L < \infty$. As will be seen below, $\alpha = 3/2$ is the relevant case. In principle, the $x$'s can be doubly subscripted, for instance, $x_1$ can change with $n$. We draw $m$ times at random without replacement from $\{x_1, \ldots, x_n\}$, generating random variables $X_1, \ldots, X_m$.

PROPOSITION A.1. *Under condition* (A31), *as* $n \to \infty$, *if* $m/n$ *converges to a positive limit that is less than* 1, *then* $\frac{1}{m}(X_1 + \cdots + X_m) - E(X_i)$ *converges in probability to* 0.

PROOF. Assume without loss of generality that $E(X_i) = 0$. Let $M$ be a positive number. Let $U_i = X_i$ when $|X_i| < M$; else, let $U_i = 0$. Let $V_i = X_i$ when $|X_i| \geq M$; else, let $V_i = 0$. Thus, $U_i + V_i = X_i$. Let $\mu = E(U_i)$, so $E(V_i) = -\mu$. Now $\frac{1}{m}(U_1 + \cdots + U_m) - \mu \to 0$. Convergence is almost sure, and rates can be given; see, for instance, Hoeffding (1963).

Consider next $\frac{1}{m}(W_1 + \cdots + W_m)$, where $W_i = V_i + \mu$. The $W_i$ are exchangeable. Fix $\beta$ with $1 < \beta < \alpha$. By Minkowski's inequality,

(A32)        $$\left[ E\left( \left| \frac{W_1 + \cdots + W_m}{m} \right|^\beta \right) \right]^{1/\beta} \leq [E(|W_i|^\beta)]^{1/\beta}.$$

When $M$ is large, the right-hand side of (A32) is uniformly small, by a standard argument starting from (A31). In essence,

$$\int_{|X_i| > M} |X_i|^\beta < M^{\beta - \alpha} \int_{|X_i| > M} |X_i|^\alpha < L/M^{\alpha - \beta}.$$

□

In proving Theorem 2, we needed $(az)_A = O(1)$. If there is an a priori bound on the absolute third moments of $a$ and $z$, then (A31) will hold for $x_i = a_i z_i$ and $\alpha = 3/2$, by the Cauchy–Schwarz inequality. On the other hand, a bound on the second moments would suffice, by Chebyshev's inequality. To get (A21) from third moments, we would, for instance, set $x_i = a_i^2$; again, $\alpha = 3/2$.

**Acknowledgments.** Donald Green generated a string of examples where the regression estimator was unbiased in finite samples; ad hoc explanations for the findings gradually evolved into Theorems 5 and 6. Sandrine Dudoit, Winston Lim, Michael Newton, Terry Speed and Peter Westfall made useful suggestions, as did an anonymous associate editor.

Department of Statistics
University of California
Berkeley, California 94720-3860
USA
E-mail: freedman@stat.berkeley.edu